\documentstyle[12pt]{article}
\begin{document}
\begin{center}
{\bf Pathway Model, Superstatistics, Tsallis Statistics, and a 
Generalized Measure of Entropy}\\[0.5cm]
{\bf A.M. Mathai}\\
Department of Mathematics and Statistics, McGill University, 805 Sherbrooke Street West, Montreal, Canada H3A 2K6\\[0.5cm]
{\bf H.J. Haubold}\\
Office for Outer Space Affairs, United Nations, Vienna International Centre, P.O. Box 500, A-1400 Vienna, Austria\\
\end{center}
\noindent
{\bf Abstract.}
The pathway model of Mathai (2005) is shown to be inferable from the maximization 
of a certain generalized entropy measure. This entropy is a variant of the 
generalized entropy of order $\alpha$, considered in Mathai and Rathie (1975), and 
it is also associated with Shannon, Boltzmann-Gibbs, R\'enyi, Tsallis, and Havrda-Charv\'at 
entropies. The generalized entropy measure introduced here is also shown to have
interesting statistical properties and it can be given probabilistic interpretations in terms 
of {\it inaccuracy measure}, {\it expected value}, and {\it information content} 
in a scheme. Particular cases of the pathway model are shown to be 
Tsallis statistics (Tsallis, 1988) and superstatistics 
introduced by Beck and Cohen (2003). The pathway model's 
connection to fractional calculus is illustrated by considering a fractional 
reaction equation.
\smallskip
\noindent
\section{Introduction}
\noindent
The fundamental problem pursued in equilibrium statistical mechanics is that given a large number of physical species, such as atoms, one wishes to know how they distribute according to some common property, e.g. velocity or energy (Ebeling and Sokolov, 2005). A simple mathematical model to understand the problem is at the center of statistics and probability theory. In order to deal with applications to physical situations of interest one takes into consideration the fundamental hypothesis of equal a priori probabilities for regions in phase space of an isolated system. This hypothesis is based on our insufficient knowledge for a specification of the precise state of the physical system under consideration. This hypothesis allows us to assign systems to states that agree equally well with our knowledge of the actual condition of the system. This leads to the Boltzmann-Gibbs entropy, or Boltzmann principle as Einstein called it, $S = k\;ln W$, where $W$ is the thermodynamic probability which is defined as the total number of equally probable microstates corresponding to the given macrostate. The Boltzmann constant is denoted by $k$. The Boltzmann-Gibbs entropy is relevant for situations such that all possible states of the system are considered equally probable. If we consider such a system in contact with a thermostat then we obtain the usual Maxwell-Boltzmann distribution for the possible states by maximizing the Boltzmann-Gibbs entropy $S$ with the normalization and energy constraints. However, in nature many systems show distributions which differ from the Maxwell-Boltzmann distribution. These are usually systems with strong autocorrelations preventing the convergence to the Maxwell-Boltzmann distribution in the sense of the central-limit theorem. Well known examples in physics are: self gravitating systems, charged plasmas, Brownian particles in the presence of driving forces, and, more generally, non-equilibrium states of physical systems (Abe and Okamoto, 2001; Gell-Mann and Tsallis, 2004). Then it is natural to ask the question of whether non-Maxwell-Boltzmannian distributions can also be obtained from a corresponding maximum entropy principle, considering a generalized form for the entropy. For this purpose, different forms were proposed, as for instance the Tsallis entropy $S_q = \frac{W^{1-q}-1}{1-q}$, where $q$ is the entropic index, that is considered the basis for a generalization of Boltzmann-Gibbs statistical mechanics (Abe and Okamoto, 2001; Gell-Mann and Tsallis, 2004). In the present paper we are investigating the link between entropic functionals and the corresponding families of distributions in Mathai's pathway model. We come to the conclusion that this link is also important to physically analyze fractional reaction equations in terms of probability theory.\\

The structure of the paper is the following: In Section 2 we introduce basic notions of Mathai's pathway model in terms of parametric families of distributions. In Section 3 we introduce a generalized entropic measure and investigate its characteristics and establish the link to parametric families of distributions in Mathai's pathway model, including Tsallis' distribution. In Section 4 we establish the link between a fractional reaction equation, its reaction coefficient considered a random variable, and Tsallis statistics and superstatistics.
\smallskip
\noindent
\section{ Preliminaries for Mathai's pathway model}
\noindent
For practical purposes of analysing data of physical experiments and in building up models in statistical physics, we 
frequently select a member from a parametric family of distributions. It is often 
found that fitting experimental data needs a model with a thicker or thinner tail than the
ones available from the parametric family, or a 
situation of right tail cut off (Honerkamp, 1994). The experimental data reveal that the underlying 
distribution is in between two parametric families of distributions. This observation either apeals to the form of the entropic functional or to the representation by a distribution function. In order 
to create a pathway from one functional form to another a pathway 
parameter is introduced and a pathway model is created in Mathai (2005). This 
model enables one to proceed from a generalized type-1 beta model to a generalized 
type-2 beta model to a generalized gamma model when the variable is restricted 
to be positive. More families are available when the variable is allowed to vary
 over the real line. Mathai (2005) deals mainly with rectangular matrix-variate 
distributions and the scalar case is a particular case there. For the real 
scalar case the pathway model is the following:
\begin{equation}
f(x)=c x^{\gamma -1}[1-a(1-\alpha)x^{\delta}]^{\frac{1}{1-\alpha}},
\end{equation}
$a>0,\delta >0,1-a(1-\alpha)x^{\delta}>0,\gamma >0$ where $c$ is the 
normalizing constant and $\alpha$ is the pathway parameter. For $\alpha <1$ the 
model remains as a generalized type-1 beta model in the real case. For 
$a=1,\gamma=1,\delta=1$ we have Tsallis statistics for $\alpha<1$ (Tsallis, 1988, 2004). Other cases 
available are the regular type-1 beta density, Pareto density, power function, 
triangular and related models. Observe that $(1)$ is a model with the right 
tail cut off. When $\alpha >1$ we may write $1-\alpha=-(\alpha -1), \alpha >1$ 
so that $f(x)$ assumes the form,
\begin{equation}
f(x)=c x^{\gamma-1}[1+a(\alpha-1)x^{\delta}]^{-\frac{1}{\alpha-1}},\;\; x>0
\end{equation}
which is a generalized type-2 beta model for real $x$. Beck and Cohen's superstatistics 
belong to this case $(2)$ (Beck and Cohen, 2003; Beck, 2006). For $\gamma=1,a=1,\delta=1$ we have Tsallis 
statistics for $\alpha>1$ from $(2)$. Other standard distributions coming from 
this model are the regular type-2 beta, the $F$-distribution, L\'evi models and 
related models. When $\alpha\rightarrow 1$ the forms in $(1)$ and $(2)$ reduce 
to
\begin{equation}
f(x)=c x^{\gamma-1} e^{- ax^\delta},\;\;x>0.
\end{equation}
\noindent
This includes generalized gamma, gamma, exponential, chisquare, Weibull, 
Maxwell-Boltzmann, Rayleigh, and related models (Mathai, 1993a; Honerkamp, 1994). If $x$ is replaced by $|x|$ in
 $(1)$ then more families of distributions  are covered in $(1)$. The 
normalizing constant $c$ for the three cases are available by putting 
$u=a(1-\alpha)x^{\delta}$ for $\alpha<1,\;u=a(\alpha-1)x^{\delta}$ for 
$\alpha>1,\;u=ax^{\delta}$ for $\alpha\rightarrow 1$ and then integrating with 
the help of a type-1 beta integral, type-2 beta integral and gamma integral 
respectively. The value of $c$ is the following:
\begin{eqnarray}
c&=& \frac{\delta [a(1-\alpha)]^{\frac{\gamma}{\delta}}
\Gamma\left(\frac{\gamma}{\delta}+\frac{1}{1-\alpha}+1\right)} 
{\Gamma\left(\frac{\gamma}{\delta}\right)\Gamma\left(\frac{1}{1-\alpha}
+1\right)},\;\;\mbox{for}\;\;\alpha<1\nonumber\\
&=&\frac{\delta [a(\alpha-1)]^{\frac{\gamma}{\delta}}\Gamma\left(\frac{1}{\alpha-1}\right)}
{\Gamma\left(\frac{\gamma}{\delta}\right)\Gamma\left(\frac{1}{\alpha-1}
-\frac{\gamma}{\delta}\right)},\;\;\mbox{for}\;\; \frac{1}{\alpha-1}
-\frac{\gamma}{\delta} >0,\;\;\alpha>1\nonumber\\
&=& \frac{\delta\;a^{\frac{\gamma}{\delta}}}{\Gamma\left(\frac{\gamma}
{\delta}\right)},\;\;\mbox {for}\;\;\alpha\rightarrow 1. 
\end{eqnarray}
Observe that in $(2)$ and $(3)$, $\frac{1}{x}$ also belongs to the same 
family of densities and hence in $(2)$ and $(3)$ one could have also taken 
$x^{-\delta}$ with $\delta>0$.
\section{ Pathway model from a generalized entropy measure}
\noindent
We introduce a generalized entropy measure here. This is a generalization of 
Shannon entropy and it is also a variant of the generalized entropy of order 
$\alpha$ in Mathai and Rathie (1975, 1976). Let us take the discrete case first. 
Consider a multinomial population $P=(p_1,...,p_k), p_i\geq 0,\; i=1,\ldots, k,\; 
p_1+\ldots +p_k=1$. Define the function
\begin{eqnarray}
M_{k,\alpha}(P)&=&\frac{\sum_{i=1}^{k}p_i^{2-\alpha}-1}
{\alpha-1},\;\;\alpha\neq 1,\;\;-\infty<\alpha<2 \\
\lim_{\alpha\rightarrow 1}M_{k,\alpha}(P)&=&-\sum_{i=1}^{k}p_i\ln p_i
=S_k(P)
\end{eqnarray}
by using L'Hospital's rule. In this notation $0\ln 0$ is taken as zero when 
any $p_i=0$. Thus $(5)$ is a generalization of Shannon entropy $S_k(P)$ as seen 
from $(6)$. Note that $(5)$ is a variant of Havrda-Charv\'at entropy 
$H_{k,\alpha}(P)$ and Tsallis entropy $T_{k,\alpha}(P)$ where
\begin{equation}
H_{k,\alpha}(P)=\frac{\sum_{i=1}^k p_i^\alpha-1}{2^{1-\alpha}-1},\;\;\alpha\neq 1,\;\;\alpha>0\\
\end{equation}
and
\begin{equation}
T_{k,\alpha}(P)=\frac{\sum_{i=1}^k p_i^\alpha-1}{1-\alpha},\;\;\alpha\neq 1,\;\;\alpha>0.
\end{equation}
We will introduce another measure associated with $(5)$ and parallel to 
R\'enyi entropy $R_{k,\alpha}$ in the following form:
\begin{equation}
M_{k,\alpha}^*(P)=\frac{\ln\left(\sum_{i=1}^k p_i^{2-\alpha}\right)}{\alpha-1},\;\;\alpha\neq 1,-\infty<\alpha<2.
\end{equation}
R\'enyi entropy is given by
\begin{equation}
R_{k,\alpha}(P)=\frac{\ln\left(\sum_{i=1}^k p_i^{\alpha}\right)}{1-\alpha},\;\;\alpha\neq 1,\;\;\alpha>0.
\end{equation}
It will be seen later that the form in $(5)$ is amenable to power law, pathway
 model etc. First we look into some basic properties enjoyed by $M_{k,\alpha}(P)$.
\subsection{Properties}
\noindent
\begin{itemize}
\item[(i)] Non-negativity: $M_{k,\alpha}(P)\geq 0$ with equality only when one 
$p_i=1$ and the rest zeros.
\item[(ii)] Expansibility or zero-indifferent: $M_{k+1,\alpha}(P,0)=M_{k,\alpha}(P)$. 
If an impossible event is incorporated into the scheme, that is, $p_{k+1}=0$ it 
will 
not change the value of the entropy measure.
\item[(iii)] Symmetry: $M_{k,\alpha}(P)$ is a symmetric function of $p_1,\ldots, p_k$. 
Arbitrary permutations of 
$p_1,\ldots,p_k$ will not alter the value of $M_{k,\alpha}(P)$.
\item[(iv)] Continuity: $M_{k,\alpha}(P)$ is a continuous function of $p_i>0,\; i=1,\ldots,k$.
\item[(v)]
Monotonicity: $M_{k,\alpha}\left(\frac{1}{k},\dots,\frac{1}{k}\right)$ is 
a monotonic increasing function of $k$.
\item[(vi)]
Inequality: $M_{k,\alpha}(p_1,\ldots,p_k)\leq 
M_{k,\alpha}\left(\frac{1}{k},\ldots,\frac{1}{k}\right).$
\item[(vii)]
Branching principle or recursivity:
\begin{eqnarray*}
M_{k,\alpha}(p_1,\ldots,p_k)&=&M_{k-1,\alpha}(p_1+p_2,p_3,\ldots,p_k)
+(p_1+p_2)^{2-\alpha}\times\nonumber\\
&\times&M_{2,\alpha}\left(\frac{p_1}{p_1+p_2},\frac{p_2}{p_1+p_2}\right).
\end{eqnarray*}
This property indicates what happens to the measure if two of the mutually 
exclusive and totally exhaustive events, defining the multinomial population, 
are combined.
\item [(viii)] Non-additivity: Consider independent multinomial populations 
$P=(p_1,\ldots, p_n)$ and $Q=(q_1,\ldots, q_m)$ such that $\sum_{i=1}^n
\sum_{j=1}^mp_iq_j=1$, $\sum_{i=1}^np_i=1$, $\sum_{j=1}^mq_j=1$. Then the joint 
density is of the form $(p_1q_1,\dots,$
$p_1q_m,
\ldots, p_nq_1,\ldots,p_n q_m)$. Let us 
denote the entropy measure in this joint distribution by $M_{nm,\alpha}(P,Q)$. 
Then
$$M_{nm,\alpha}(P,Q)=M_{n,\alpha}(P)
+M_{m,\alpha}(Q)+(\alpha-1)M_{n,\alpha}(P)M_{m,\alpha}(Q).
$$
The third term on the right makes the measure non-additive. But the measures 
$M_{k,\alpha}^{*}(P)$ and R\'enyi entropy $R_{k,\alpha}(P)$ as well as Shannon 
entropy $S_k$ are all additive. This is due to the fact that when logarithm of 
a product is taken it leads to a sum.
\item[(ix)] Decomposibility: Consider a joint discrete distribution\\
 $p_{ij}\geq 0$, 
$\sum_{i=1}^n\sum_{j=1}^mp_{ij}=1$. Consider the marginal distribution 
$P_j=\sum_{i=1}^np_{ij}>0,\;j=1,\ldots, m$. Then we have
\begin{eqnarray*}
&&M_{nm,\alpha}(p_{11},p_{12},\ldots, p_{1m},p_{21},\ldots, p_{2m},\ldots,
p_{n1},\ldots, p_{nm})\\
& =& M_{m,\alpha}(P_1,\ldots, P_m)
+\sum_{j=1}^m P_j^{2-\alpha} M_{n,\alpha}\left(\frac{p_{1j}}{P_j},\dots,\frac{p_{nj}}{P_j}\right).
\end{eqnarray*}
Observe that when $p_{ij}=p_iq_j$ with $\sum_{i=1}^np_i=1$, $\sum_{j=1}^mq_j=1$ 
then property $(ix)$ reduces to  property $(viii)$.
\item[(x)]
Functional equation: Consider $M_{2,\alpha}(P)=M_{2,\alpha}(p,1-p)$. 
That is,
$$M_{2,\alpha}(p,1-p)=\frac{p^{2-\alpha}+(1-p)^{2-\alpha}-1}{\alpha-1},
\;\;\alpha\neq 1,\;\alpha<2.$$
Let $f_{\alpha}(p)=M_{2,\alpha}(p,1-p)$. Then $f_{\alpha}(p)$ satisfies the 
functional equation
\begin{equation}
f_\alpha(x)+(1-x)^{2-\alpha}f_\alpha \left(\frac{y}{1-x}\right)
=f_\alpha(y)+(1-y)^{2-\alpha}f_\alpha \left(\frac{x}{1-y}\right)
\end{equation}
for $x,y\in [0,1), x+y\in [0,1]$ where $[\cdot )$ means open on the right and 
closed on the left, with
\begin{equation}
f_\alpha (0)=f_\alpha(1)=0 
\end{equation}
and
\begin{equation}
f_\alpha\left(\frac{1}{2}\right)=\frac{2^{\alpha-1}-1}{\alpha-1},\;\;\alpha\neq 1.
\end{equation}
\end{itemize}
\smallskip
\noindent
\subsection{ Characterization}
\smallskip
Several characterization theorems can be established for the measure of generalized 
entropy in $(5)$. We will indicate one such characterization. Let us look at an 
arbitrary continuous function, call it $f_\alpha(x)$, satisfying the 
functional equation in $(11)$ with the boundary conditions in $(12)$ and $(13)$.
 What is the functional form of $f_{\alpha}(x)$? We will solve for 
$f_\alpha(x)$ from $(11)$. Put $y=1-x$ in $(11)$ and use the boundary condition $(12)$ to obtain
$$f_\alpha(x)=f_\alpha(1-x).$$
Take two numbers $p$ and $q$ in the open interval $(0,1)$. Put $p=1-x$ and $q=y/(1-x)$ in $(11)$. Then we obtain
$$f_\alpha(p)+p^{2-\alpha}f_\alpha(q)=f_\alpha(pq)+(1-pq)^{2-\alpha}f_\alpha\left(\frac{1-p}{1-pq}\right).$$
Let
$$F(p,q)=f_\alpha(p)+[p^{2-\alpha}+(1-p)^{2-\alpha}]f_\alpha(q).$$
Then it is easily seen that $F(p,q)=F(q,p)$, which implies that
$$f_\alpha(p)+[p^{2-\alpha}+(1-p)^{2-\alpha}]f_\alpha(q)
=f_\alpha(q)
+[q^{2-\alpha}+(1-q)^{2-\alpha}]f_\alpha(p).$$
Put $q=1/2$ in this equation and use the boundary condition $(13)$ to see that
$$f_\alpha(p)=\frac{p^{2-\alpha}+(1-p)^{2-\alpha}-1}{\alpha-1}.$$
Now we can build up successively by using the branching property $(viii)$.\\
{\bf Theorem 2.1.} {\it Let $f_{k}(p_1,...,p_k)$ be an arbitrary 
function of $p_1,\ldots, p_k,\;\;p_i\geq 0,\;\;i=1,\ldots,k,\;\;p_1+\ldots+p_k=1$ satisfying 
the properties of symmetry (property $(iii)$ of section 3.1), continuity 
(property $(iv)$ of section 3.1), branching principle (property $(vii)$ of 
section 3.1) where $f_2(p,1-p)=f_\alpha(p)$, satisfies the functional equation
 $(11)$ with the boundary conditions in $(12)$ and $(13)$ then $f_k(p_1,\ldots p_k)$
 is uniquely determined as $M_{k,\alpha}(P)$ of $(5)$.}
\vskip.2cm
\noindent
\subsection{Continuous analogue}
\smallskip
The continuous analogue to the measure in $(5)$ is the following:
\begin{eqnarray}
M_\alpha(f)&=&\frac{\int_{-\infty}^\infty[f(x)]^{2-\alpha}dx-1}
{\alpha-1}\\
&=&\frac{\int_{-\infty}^\infty[f(x)]^{1-\alpha}f(x)dx-1}
{\alpha-1}=\frac{E[f(x)]^{1-\alpha}-1}{\alpha-1},\;\;\alpha\neq 1,\;\;\alpha <2\nonumber
\end{eqnarray}
where $E[\cdot]$ denotes the expected value of $[\cdot]$. 
Note that when $\alpha =1,\;\;E[f(x)]^{1-\alpha}\\
=E[f(x)]^0=1.$\\
When $\alpha <0$ and decreases then $1-\alpha>1$ and increases. The measure of uncertainty
decreases in the discrete case when 
$\alpha<0$. Similarly when $\alpha>0$, then $1-\alpha<1$ and decreases. In the 
discrete case the measure of uncertainty 
increases. Hence we may call $1-\alpha$ as the {\it strength of information} in 
the distribution. Larger the value of $1-\alpha$ the larger the information 
content and smaller the uncertainlty and vice versa.
\smallskip
\noindent
\subsection{Connection to Kerridge's ``inaccuracy'' measure}
A connection to Kerridge's measure of ``inaccuracy'' can also be explored here. 
Kerridge (1961) defined
\begin{equation}
K(P,Q)=-\sum_{i=1}^n p_i\ln q_i,
\end{equation}
$P=(p_1,\ldots, p_n),\;\;Q=(q_1,\ldots, q_n),\; p_i\geq 0,\;\;q_i\geq 0,\;\; i=1,\ldots, n$, 
$\sum_{i=1}^n p_i=1,\sum_{i=1}^n q_i=1$, as a measure of ``inaccuracy'' in assigning
 the multinomial distribution $Q$ for the true distribution $P$. Here $Q$ may 
be relative frequencies coming from an experiment or observations and $P$ may 
be the true underlying distribution. A generalization of $(15)$ can be the following:
\begin{equation}
M_\alpha(P,Q)=\frac{\sum_{i=1}^n p_i q_i^{1-\alpha}-1}{\alpha-1},\;\;\alpha\neq 1,
\end{equation}
so that when $\alpha\rightarrow 1$ then $M_\alpha (P,Q)$ goes to $K(P,Q)$ of 
$(15)$. Observe that we can also look upon $M_\alpha (P,Q)$ as an expected value 
with respect to the true multinomial distribution $P$ and we may write it as
\begin{equation}
M_\alpha (P,Q)=\frac{E[q^{1-\alpha}]-1}{\alpha-1},\;\;\alpha\neq 1,
\end{equation}
$q$ taking the values $q_1,\ldots, q_n$ with the corresponding probabilities 
$p_1,\ldots, p_n$. The continuous analogue of $(17)$ is the following:
\begin{equation}
M_\alpha(f_1,f_2)
=\frac{E[f_2^{1-\alpha}]-1}{\alpha-1}
=\frac{\int_{-\infty}^\infty f_1(x)[f_2(x)]^{1-\alpha}dx-1}
{\alpha-1},\;\;\alpha\neq 1
\end{equation}
where the expectation is taken with respect to the density $f_1(x)$. Instead of
$[f_2(x)]^{1-\alpha}$ if we assign $[f_1(x)]^{1-\alpha}$ through a displacement
 or distortion or disturbance to $f_1(x)$ then a measure of ``inaccuracy''in 
assigning $[f(x)]^{1-\alpha}$ for $f(x)$ is given by $M_\alpha(f)$ of $(14)$. 
More ``inaccuracy'' means less ``information content'' and more ``uncertainty''or 
more  
``entropy'', and vice versa. Again $1-\alpha$ corresponds to the strength of 
information content.\\

The generalized entropy $M_\alpha(f)$ is evaluated for some standard distributions 
and given in the appendix at the end of this paper.
\smallskip
\noindent
\subsection{Distributions with maximum generalized entropy}
Among all densities, which one will give a maximum value for $M_{\alpha}(f)$ ? 
Consider all possible functions $f(x)$ such that $f(x)\geq 0$ for all $x$, 
$f(x)=0$ outside $(a,b)$, $a<b$, $f(a)$ is the same for all such $f(x)$, 
$f(b)$ is the same for all such $f$, $\int_{a}^b f(x)dx=1$. Let $f(x)$ 
be a continuous function of $x$ possessing continuous derivatives with respect 
to $x$. Then for using calculus of variation techniques consider
\begin{equation}
U=[f(x)]^{2-\alpha}-\lambda\;\;f(x).
\end{equation}
Note that for fixed $\alpha,\alpha\neq 1$, maximization of 
$\frac{\int_a^b [f(x)]^{2-\alpha}dx-1}{\alpha-1},\;\;\alpha\neq 1,
\alpha<2$ is equivalent to maximizing $\int_a^b[f(x)]^{2-\alpha}dx$. 
If necessary, we may also take
$$M_\alpha(f)=\frac{\int_a^b[f(x)]^{2-\alpha}dx}{\alpha-1}-\frac{\int_a^b f(x)dx}{\alpha-1},\;\;\alpha\neq 1,\;\;\alpha<2$$
since $\int_a^b f(x)dx=1$. This will produce only a change in the 
Lagrangian multiplier $\lambda$ in $U$ above. Hence without loss of generality 
the form of $U$ is as given in $(19)$. We are looking at all possible $f$ for 
every given $x$ and $\alpha$. Hence the Euler equation becomes,
\begin{eqnarray*}
\frac{\partial U}{\partial f}=0 &\Rightarrow& (2-\alpha)[f(x)]^{1-\alpha}-\lambda=0 \\
&\Rightarrow& f(x)=\frac{\lambda}{2-\alpha},
\end{eqnarray*}
free of $x$, $\alpha<2,\;\;\alpha\neq 1$. Thus $f(x)$ in this case is a uniform density over
$[a,b]$.\\
Let us consider the situation where $E[x^{\delta}]$ for some $\delta$ is a fixed quantity for all such $f$. Then we have to maximize
$$\frac{\int_a^b [f(x)]^{2-\alpha}dx}{\alpha-1}-\frac{1}{\alpha-1}$$
subject to the conditions $\int_a^b f(x)dx=1$ and $\int_a^b x^\delta f(x)dx$ is a given quantity. Consider
$$U=[f(x)]^{2-\alpha}-\lambda_1 f(x)+\lambda_2 x^\delta f(x).$$
Then the Euler equation is the following:
\begin{eqnarray}
\frac{\partial U}{\partial f} =0 &\Rightarrow& (2-\alpha)[f(x)]^{1-\alpha}-\lambda_1+\lambda_2x^\delta=0\nonumber\\
&\Rightarrow& [f(x)]^{1-\alpha}=\frac{\lambda_1}{2-\alpha}[1-\frac{\lambda_2}{\lambda_1}x^\delta]\nonumber\\
&\Rightarrow& f(x)=c_1[1-c_2x^\delta]^{\frac{1}{1-\alpha}}
\end{eqnarray}
where $c_1$ and $c_2$ are constants and $c_1>0,1-c_2x^\delta>0$ since it is 
assumed that $f(x)\geq 0$ for all $x$. When $c_2=\beta(1-\alpha),\;\;\beta>0$, we 
have
\begin{equation}
f(x)=c_1[1-\beta(1-\alpha)x^{\delta}]^{\frac{1}{1-\alpha}}.
\end{equation}
Then for $\delta=1$ we have the power law
\begin{equation}
\frac{\partial f}{\partial x}=-c_3f^\alpha
\end{equation}
where $c_3$ is a constant. The form in $(21)$ for $\alpha<1$ remains as a 
special case of a generalized type-1 beta model; for $\alpha >1$ it is a special 
 case of a generalized type-2 beta 
model and when $\alpha\rightarrow 1$ it is a special case of a generalized 
gamma model when the 
range $(a,b)$ is such that $a=0,b=\infty$. For $\delta=1$, $(21)$ gives Tsallis 
statistics (Tsallis, 1988, 2004).\\
Observe that the generalized entropy $M_\alpha(f)$ of $(14)$ gives rise to the
power law with exponent $\alpha$, readily, as seen from $(22)$. Also notice that
by selecting $\lambda_1$ and $\lambda_2$ in $(20)$ we can obtain functions of 
the following forms also:
$$(1-\beta_1 x^\delta)^{-\gamma_1}\;\mbox{and}\;(1+\beta_2 x^\delta)^{\gamma_2},\;\;\beta_1,\beta_2,\gamma_1,\gamma_2>0.
$$
Both these forms are ever increasing and cannot produce densities in 
$(0,\infty)$ unless the range of $x$ with nonzero $f(x)$ is finite.\\
In section 3.4 we have given several interpretations for $1-\alpha$. We can also
derive the pathway model by maximizing $M_{\alpha}(f)$ over all non-negative 
integrable functions. Consider all possible $f(x)$ such that $f(x)\geq 0$ for all
$x$, $\int_a^b f(x)dx<\infty$, $f(x)$ is zero outside $(a,b)$, $f(a)$ 
is the same for all $f(x)$, and similarly $f(b)$ is also the same for all such 
functional $f$. Let $f(x)$ be a continuous function of $x$ with continuous 
derivatives in $(a,b)$. Let us maximize $\int_a^b [f(x)]^{2-\alpha}dx$ 
for fixed $\alpha$ and over all functional $f$, under the conditions that the 
following two moment-like expressions be fixed quantities:
\begin{equation}
\int_a^b x^{(\gamma-1)(1-\alpha)}f(x)dx=\mbox{given, and}\;
\int_a^b x^{(\gamma-1)(1-\alpha)+\delta}f(x)dx=\mbox{given}
\end{equation}
for fixed $\gamma>0$ and $\delta>0$. Consider
$$U=[f(x)]^{2-\alpha}-\lambda_1 x^{(\gamma-1)(1-\alpha)}f(x)+\lambda_2 x^{(\gamma-1)(1-\alpha)+\delta} f(x),\;\;\alpha<2,\;\;\alpha\neq 1$$
where $\lambda_1$ and $\lambda_2$ are Lagrangian multipliers. Then the Euler equation is the following:
\begin{eqnarray}
\frac{\partial U}{\partial f} =0 &\Rightarrow& (2-\alpha)[f(x)]^{1-\alpha}-\lambda_1x^{(\gamma-1)(1-\alpha)}+\lambda_2x^{(\gamma-1)(1-\alpha)+\delta}=0\nonumber\\
&\Rightarrow& [f(x)]^{1-\alpha}=\frac{\lambda_1}{(2-\alpha)}x^{(\gamma-1)(1-\alpha)}[1-\frac{\lambda_2}{\lambda_1}x^\delta]\\
&\Rightarrow& f(x)=c_1x^{\gamma-1}[1-\beta(1-\alpha)x^\delta]^{\frac{1}{1-\alpha}}
\end{eqnarray}
where $\lambda_1/\lambda_2$ is written as $\beta(1-\alpha)$ with $\beta>0$ 
such that $1-\beta (1-\alpha)x^\delta>0$ since $f(x)$ is assumed to be 
non-negative. By using the conditions $(23)$ we can determine $c_1$ and $\beta$.
 When the range of $x$ for which $f(x)$ is nonzero is $(0,\infty)$ and when 
$c_1$ is a normalizing constant then $(25)$ is the pathway model of Mathai 
(2005) in the scalar case where $\alpha$ is the pathway parameter. When 
$\gamma=1,\delta=1$ then $(25)$ produces the power law. The form in $(24)$ for 
various values of $\lambda_1$ and $\lambda_2$ can produce all the four forms
$$\alpha_1 x^{\gamma-1}[1-\beta_1(1-\alpha)x^\delta]^{-\frac{1}{1-\alpha}},\;
\alpha_2 x^{\gamma-1}[1-\beta_2(1-\alpha)x^\delta]^{\frac{1}{1-\alpha}}\;\mbox{for}\;\alpha<1$$
\noindent
and
$$\alpha_3 x^{\gamma-1}[1+\beta_3(\alpha-1)x^\delta]^{-\frac{1}{\alpha-1}},\;
\alpha_4 x^{\gamma-1}[1+\beta_4(\alpha-1)x^\delta]^{\frac{1}{\alpha-1}}\;\mbox{for}\;\alpha>1$$
with $\alpha_i,\beta_i>0,i=1,2,3,4$. But out of these, the second and the third
 forms can produce densities in $(0,\infty)$. The first and fourth will not be 
converging. When $f(x)$ is a density in $(25)$ what is the normalizing constant 
$c_1$? We need to consider three cases of $\alpha<1,\alpha>1$ and 
$\alpha\rightarrow 1$. This $c_1$ is already evaluated in section 2.
\smallskip
\noindent
\section{Reaction equation, superstatistics, and fractional calculus}
\noindent
\subsection {Reaction equation and fractional calculus}
A reaction problem was examined by Haubold and Mathai (1995, 2000) 
where the number density of the reacting particles is a function of time $t$. 
For the $i$-th particle let the number density be denoted by $N_i(t)$ with 
$N_i(t=0)=N_0^{(i)}$. If the production rate is proportional to the number density 
then we have
$$\frac{d}{dt}N_i(t)=a_i N_i(t)$$
where $a_i$ is the reaction coefficient. The reaction coefficient itself can be considered to be a statistical quantity subject to accommodating a distribution function based either on Boltzmann-Gibbs or Tsallis statistics (Haubold and Mathai, 1998; Coraddu at al., 1999; Saxena, Mathai, and Haubold 2004b). If some of the particles produced are also 
destroyed and if the destruction rate is $b_i$ for the $i$-th 
particle, and proportional to the number density, then
$$\frac{d}{dt} N_i(t)=b_iN_i(t).$$
If destruction dominates then the net residual effect is given by
$$\frac{d}{dt} N_i(t)=-c_iN_i(t), ~c_i=b_i-a_i>0$$
and the solution is
$$N_i(t)=N_0^{(i)}e^{-c_it},\;\;c_i>0, t\geq 0.$$
But if a fractional integral, instead of the integer-order integral, is used then the 
residual reaction equation is given by
\begin{equation}
N_i(t)=N_0^{(i)}-c_i^{\nu}\;\;_0D_t^{-\nu} N_i(t), \;\;\nu>0
\end{equation}
where the fractional integral operator is given by
\begin{equation}
_aD_t^{-\nu} f(t)=\frac{1}{\Gamma(\nu)}\int_a ^t (t-u)^{\nu-1}f(u) du,\;\;\nu>0,
\end{equation}
with $_aD_t^0 f(t)=f(t)$. Note that in $(26)$, $c_i$ is replaced by 
$c_i^\nu$ for physical reason. If the production rate dominates over the 
destruction rate then in $(26)$, $c_i^\nu$ is to be replaced by $-c_i^\nu$. 
For a discussion of general growth-decay models see Mathai (1993b). The Laplace 
transform of $N_i(t)$ coming from $(26)$ is
\begin{equation}
L_{N_i(t)}(s)=\frac{N_0^{(i)}}{s\left[1+\left(\frac{c_i}{s}\right)^\nu\right]}
\end{equation}
and the solution can be written in terms of a Mittag-Leffler function:
\begin{equation}
N_i(t)=N_0^{(i)}\sum_{k=0}^\infty \frac{(-1)^k[(c_it)^\nu]^k}
{\Gamma(1+k\nu)}= N_0^{(i)}\;\; E_\nu(-c_i^\nu t^\nu)
\end{equation}
where the generalized Mittag-Leffler function $E_{\mu,\nu}^{\gamma}(z)$ is 
defined by
\begin{equation}
E_{\alpha,\beta}^{\gamma}(z)=\sum_{k=0}^\infty \frac{(\gamma)_k\;z^k}
{k!\Gamma(\beta+\alpha k)},\;\;\beta>0,\;\;\alpha >0
\end{equation}
 where $(\gamma)_k=\gamma(\gamma+1)...(\gamma+k-1),\;\;(\gamma)_0=1,\;\;\gamma\neq 0$ 
and 
$$E_{\alpha,1}^1(z)=E_\alpha(z)=\sum_{k=0}^\infty \frac{z^k}
{\Gamma(1+\alpha k)},\;\;\alpha > 0.$$
Consider the situation of having a total of $\gamma$ different particles, 
$\gamma=1,2,\ldots$, these particles reacting independently with the same rate 
$c_1=c_2=\ldots = c_\gamma = c$. For independent variables the Laplace transform of 
the sum is the product of Laplace transforms. Hence in this case 
$N(t)=N_1(t)+...+N_\gamma(t)$ has the Laplace transform
\begin{equation}
L_{N(t)}(s)=\frac{N_0^{(1)}\ldots N_0^(\gamma)}{s^\gamma\left[1+\left(\frac{c}{s}\right)^\nu\right]^\gamma}
\end{equation}
which yields the solution
\begin{equation}
N(t)=N_0^{(1)}\ldots N_0^{(\gamma)}\;\;t^{\gamma-1}E_{\nu,\gamma}^\gamma(-t^\nu c^\nu).
\end{equation}
The rate $c$ may vary and $c$ may be considered to have its own distribution. In this case 
$N(t)$ in $(32)$ is to be taken as $N(t|c)$ or $N(t)$ at a given $c$. 
Superstatistics, as considered by Beck and Cohen (2003) and Beck (2006), will result when we use an appropriate prior density for $c$ as developed in the following section.
\smallskip
\noindent
\subsection{Superstatistics and fractional calculus}
Beck and Cohen developed the concept of superstatistics, 
see Beck and Cohen (2003) and Beck (2006). They showed that superstatistics considerations are 
relevant in problems of Lagrangian and Eulerian turbulence, defect turbulence, 
atmospheric turbulence, cosmic ray statistics, solar flares, and solar wind 
statistics. From a statistical point of view, the 
procedure is equivalent to starting with a conditional distribution for every 
given value of a parameter $\beta$. Then $\beta$ is assumed to have a prior 
known density. Then the unconditional distribution is obtained by integrating 
out over the density of $\beta$.\\

Reaction equations are considered in a series of papers by Haubold, Mathai, and Saxena (Haubold and Mathai, 1995, 2000; Saxena, Mathai, and Haubold, 2004a; Mathai, Saxena, and Haubold, 2005). For the 
sake of illustration we consider one such reaction equation which 
yields the number density of the following form, where $\gamma$ and $\mu$ are 
arbitray, need not be integers. [Note that in $(32)$ the resulting $\gamma$ 
is a positive integer.] 
\begin{eqnarray}
N(t|c)&=& N_0\;t^{\mu-1}E_{\nu,\mu}^{\gamma+1}(-c^\nu t^\nu),\;~\mu>0,\;\gamma>0,\;\nu>0\nonumber\\
&=& N_0\; t^{\mu-1}\sum_{k=0}^\infty \frac{(\gamma+1)_k(-1)^k(c^\nu t^\nu)^k}{k!\;\Gamma(\mu+\nu k)}.
\end{eqnarray}
Let us consider the situation where $c$ in equation (26) is a random variable having a gamma 
type density:
\begin{equation}
g(c)=\frac{\omega^\mu}{\Gamma(\mu)}c^{\mu-1}e^{-\omega c},\;\omega>0,\; 0<c<\infty
\end{equation}
where $\mu>0,\omega,\mu$ are known and $\mu/\omega$ is the mean value 
of $c$. The residual rate of change may have small probabilities of it being too
 large or too small and the maximum probability may be for a medium range of 
values for the residual rate of change $c$. This is a very reasonable asumption.
 Equation $(33)$ is the situation where the residual rate of change is such that 
the production rate dominates so that we have the form $-c^\nu,\nu>0,c>0$. If the destruction 
rate dominates then the 
constant will be of the form $c^\nu, c>0,\nu>0$. Integrating out $N(t|c)$ over 
$g(c)$ we have
$$\int_{c=0}^\infty N(t|c)g(c)dc=N_0 \frac{\omega^\mu}{\Gamma(\mu)}t^{\mu-1}\sum_{k=0}^\infty \frac{(\gamma+1)_k(-1)^k(t^{\nu})^k}{k!\Gamma(\mu+\nu k)} \int_0^\infty c^{\mu-1+k\nu} e^{-c\omega}dc$$
where the integral over $c$ yields $\Gamma(\mu+k\nu)\omega^{-(\mu+k\nu)}$. 
Hence the unconditional number density, denoted by $N(t)$, is given by
\begin{eqnarray*}
N(t)&=& \frac{N_0}{\Gamma(\mu)}t^{\mu-1}
\sum_{k=0}^\infty \frac{(\gamma+1)_k (-1)^k}{k!}\left(\frac{t^\nu}{\omega^\nu}\right)^k\\
&=& \frac{N_0}{\Gamma(\mu)}t^{\mu-1}\left(1+\frac{t^\nu}{\omega^\nu}\right)^{-(\gamma+1)}\;\mbox{for}\;\;|\frac{t}{\omega}|<1.
\end{eqnarray*}
From the analytic continuation we see that the form remains the same for 
$\frac{t}{\omega}>1$ also. Hence
\begin{equation}
N(t)=\frac{N_0}{\Gamma(\mu)}t^{\mu-1}\left(1+\frac{t^\nu}
{\omega^\nu}\right)^{-(\gamma+1)},\; 0<t<\infty,\;\omega>0.
\end{equation}
The continuation part may be seen by writing equation $(33)$ with the help of 
the Mellin-Barnes representation in the form
\begin{equation}
N(t|c)=N_0 t^{\mu-1}\frac{1}{2\pi i}\int_{b-i\infty}^{b+i\infty}
\frac{\Gamma(1-s)\Gamma(s)}{\Gamma(1-s\nu)}(c^\nu t^\nu)^{-s}ds,\; i=\sqrt{-1}.
\end{equation}
Then integrating over $g(c)$ of $(34)$ we have
\begin{equation}
N(t)=N_0 t^{\mu-1}\frac{1}{2\pi i}\int_{b-i\infty}^{b+i\infty}\Gamma(1-s)\Gamma(s)\left(\frac{t^\nu}{\omega^\nu}\right)^{-s}ds
\end{equation}
where $b$ is a real number, $0<b<1$. Evaluate the integral in $(37)$ as the 
sum of the residues at the poles of $\Gamma(s)$, which are at $s=-k,k=0,1,...$, 
to obtain the series form for ${{t}\over{\omega}}<1$. Evaluate at the poles of 
$\Gamma(1-s)$ to obtain the analytic continuation for ${{t}\over{\omega}}>1$. 
Both will lead to the same form in $(35)$, which is a generalized type-2 beta 
form.\\
We may make the substitution $\gamma+1=\frac{1}{\alpha-1},\alpha>1\Rightarrow \gamma
=\frac{\alpha-2}{\alpha-1}$ and $\omega^{-\nu}= b(\alpha-1), b>0$. Then we have 
the unconditional number density
\begin{equation}
N(t)=\frac{N_0}{\Gamma(\mu)}t^{\mu-1}[1+b(\alpha-1)t^\nu]^{-\frac{1}{\alpha-1}}
\end{equation}
for $\alpha>1,t>0,b>0,\mu>0$. For $\mu=1,\nu=1,b=1$, equation $(38)$ 
corresponds to Tsallis statistics for one case, namely, $\alpha =q>1$ (Tsallis 2004). 
For general values of $\mu$ and $\alpha >1$ such that 
$\frac{1}{\alpha-1}-\frac{\mu}{\nu}>0$ equation $(38)$ corresponds to 
the pathway model of Mathai (2005) as well as the superstatistics considered 
by Beck and Cohen (2003) and Beck (2006).\\
Now, consider an equation parallel to equation (17) in Mathai, Saxena, and Haubold (2005), namely,
\begin{equation}
N(t|c)-N_0 t^{\mu-1}E_{\nu,\mu}^{-\gamma}(c^{\nu}t^\nu)=c^\nu~{_0D_t^{-\nu}}N(t)
\end{equation}
for $\gamma>0,c>0,\nu>0,\mu>0$. Note that sets of parameters, 
$\gamma>0,\nu>0,\mu>0$ exist so that the series form in 
$E_{\nu,\mu}^{-\gamma}(c^{\nu}t^{\nu})$ is convergent and remains positive for 
all $t>0$. This corresponds to the situation where the residual rate of change 
is positive so that the production dominates over 
destruction. We can expect $N(t|c)$ to increase to $\infty$. But the density of 
$c$ in $(34)$ will produce a dampening effect. Proceeding as before, we will 
obtain the final unconditinal $N(t)$ as follows:
\begin{equation}
N(t)=\frac{N_0}{\Gamma(\mu)}t^{\mu-1}\left[1-\frac{t^\nu}{\omega^\nu}\right]^{\gamma-1}\;\; \mbox{for}\;0<\frac{t}{\omega}
<1.
\end{equation}
For convenience, we can write $\gamma-1=\frac{1}{1-\alpha},\alpha<1\Rightarrow \gamma=\frac{2-\alpha}{1-\alpha}$ and $\omega^{-\nu}= b(1-\alpha), b>0, \alpha<1$ so that $(40)$ becomes
\begin{equation}
N(t)=\frac{N_0}{\Gamma(\mu)}t^{\mu-1}[1-b(1-\alpha)t^\nu]^{\frac{1}{1-\alpha}},
\end{equation}
for $\alpha <1,b>0,\nu>0,\mu>0,0<t^{\nu}<[b(1-\alpha)]^{-1}$. If $N(t)$ is to 
be made into a  statistical density then we may normalize $(38)$ with the help of a type-2 
beta integral, and $(41)$ with the help of a type-1 beta integral. When 
$\alpha \rightarrow 1$ both $(38)$ and $(41)$ will approach the generalized gamma 
form
\begin{equation}
N(t)=\frac{N_0}{\Gamma(\mu)}t^{\mu-1}e^{-bt^{\nu}},\;b>0,\mu>0,\nu>0,t>0.
\end{equation}
Equations $(38),(41)$ and $(42)$ are special scalar cases
of the pathway model of Mathai (2005). For $\mu=1,\nu=1,b=1$ 
Tsallis statistics is recovered from $(38)$ and $(41)$, with appropriate 
normalizations.\\
{\bf Remark 1.} Superstatistics considerations can produce only the 
generalized type-2 beta form of $(35)$ with the appropriate normalization 
constant because Beck and Cohen (2003) and Beck (2006) consider the situation of a conditional 
density belonging to the generalized gamma family where a scaling parameter is 
having a marginal distribution belonging to a generalized gamma family. Then the
 unconditional density can only produce a form of the type
\begin{equation}
f(x)=c_1x^\alpha(1+\beta_1 x^\delta)^{-\gamma_1},\;c_1>0,\beta_1>0,\gamma_1>0,x>0
\end{equation}
which is a generalized type-2 beta form. A type-1 beta form as in $(40)$ will
 not be available from such a procedure because the exponents will only sum up 
to give a form of the type in $(43)$. A more general format of the Beck-Cohen type 
consideration, from a statistical point of view, is provided in the following section.

\subsection{A more general density giving rise to generalized superstatistics}
Consider a conditional density of the form
\begin{equation}
f_{x}(x|\beta)=c_2(x-x_o)^{\alpha-1}e^{-(\beta -\beta_o)^\eta(x-x_o)^\delta},
\end{equation}
for $\eta>0,\delta>0,\alpha>0,\beta>\beta_o,x>x_o$ where the normalizing 
constant $c_2$ can be seen to be the following:
\begin{equation}
c_2=\frac{\delta \;(\beta-\beta_o)^{\frac{\alpha\eta}{\delta}}}
{\Gamma\left(\frac{\alpha}{\delta}\right)}.
\end{equation}
For various values of the parameters $\alpha,\eta,\delta, x_o, \beta_o$ we have the 
exponential, gamma, chisquare, Erlang, Helley, Helmert, Maxwell-Boltzmann, 
Rayleigh, and Weibull densities as special cases of $(44)$ (Mathai, 1993a; Honerkamp, 1994). If $x-x_o$ is 
replaced by $|x-x_o|$ for $-\infty<x<\infty$ then many more densities such as 
double exponential, Laplace and Gaussian will be particular cases. Consider a 
prior density for the scaling parameter $\beta$ in the following form:
\begin{equation}
g(\beta)=c_3(\beta-\beta_o)^{\gamma-1}e^{-\zeta (\beta-\beta_o)^{\eta}},
\end{equation}
for $\zeta>0,\eta>0,\gamma>0$ known, $\beta>\beta_o$ and $c_3
=\eta\zeta^{\frac{\gamma}{\eta}}/\Gamma\left(\frac{\gamma}{\eta}\right)$. This
 is a very general class of the type in $(44)$. The only restriction imposed for
 convenience is that the exponents involving $\beta$ are the same in $(44)$ and 
$(46)$. Then the unconditional density of $x$ is given by
\begin{eqnarray}
f_{x}(x)&=&\int_{\beta =\beta_o}^{\infty}f_{x}(x|\beta)g(\beta)d\beta \nonumber\\
&=& \frac{\eta \;\delta\,\zeta^{\frac{\gamma}{\eta}}(x-x_o)^{\alpha-1}}{\Gamma\left(\frac{\gamma}{\eta}\right)\Gamma\left(\frac{\alpha}{\delta}\right)}
\int_{\beta =\beta_o}^\infty(\beta -\beta_o)^{\gamma+\frac{\alpha \eta}{\delta}-1}e^{-(\beta -\beta_o)^{\eta}[\zeta+(x-x_o)^{\delta}]}d\beta\nonumber\\
&=&\frac{\delta\;\Gamma\left(\frac{\gamma}{\eta}+\frac{\alpha}{\delta}\right)\;(x-x_o)^{\alpha -1}\left[1+\frac{(x-x_o)^{\delta}}{\zeta}\right]^{-\left(\frac{\gamma}{\eta}+\frac{\alpha}{\delta}\right)}}{\zeta^{\frac{\alpha}{\delta}}\Gamma\left(\frac{\gamma}{\eta}\right)
\Gamma\left(\frac{\alpha}{\delta}\right)}
\end{eqnarray}
for $x>x_o,\zeta>0,\gamma>0,\eta>0,\alpha>0,\delta>0$. Note that $(47)$ can 
only produce a form of the type $(43)$ or a generalized type-2 beta type. Type-1
 beta form of $(40)$ is not available from a procedure such as the steps in 
$(44)$ to $(47)$. This is not covered in Beck-Cohen's type superstatistics 
considerations.\\
{\bf Remark 2.} Observe that if a real random variable $x$ has a 
density belonging to the generalized gamma family then $\frac{1}{x}$ also 
belongs to a generalized gamma family. Hence in $(44)$ and $(46)$ we could have 
taken $(x-x_o)^{-\delta}$ with $\delta>0$ and $(\beta-\beta_o)^{-\eta}$ with 
$\eta>0$ in the exponents, individually or simultaneously. The procedure 
will remain the same and we will again 
end up with a generalized type-2 beta form corresponding to $(47)$.
\smallskip
\begin{center}
{\bf References}
\end{center}
\smallskip
\noindent
Abe, S. and Okamoto, Y. (Eds.)(2001). {\it Nonextensive Statistical Mechanics and Its Applications}. Springer-Verlag, Berlin and Heidelberg.\par
\smallskip
\noindent
Beck, C. (2006). Stretched exponentials from superstatistics. 
{\it Physica}, {\bf A 365}, 96-101.\par
\smallskip
\noindent
Beck, C. and Cohen, E.G.D. (2003). Superstatistics. {\it Physica}, {\bf A322},
267-275.\par
\smallskip
\noindent
Coraddu, M., Kaniadakis, G., Lavagno, A., Lissia, M., Mezzorani, G., and Quarati, P. (1999). Thermal distributions in stellar plasmas, nuclear reactions and solar neutrinos. {\it Braz. J. Phys.}, {\bf 29}, 153-168 [see also Lissia, M. and Quarati, P. (2005). Nuclear astrophysical plasmas: ion distribution functions and fusion rates. {\it Europhysics News}, {\bf November/December}, 211-214].\par
\smallskip
\noindent
Ebeling, W. and Sokolov, I.M. (2005). {\it Statistical Thermodynamics and Stochastic Theory of Nonequilibrium Systems}. World Scientific, Singapore.\par
\smallskip
\noindent
Gell-Mann, M. and Tsallis, C. (Eds.)(2004). {\it Nonextensive Entropy: Interdisciplinary Applications}. Oxford University Press, New York.\par
\smallskip
\noindent
Haubold, H.J. and Mathai, A.M. (1995). A heuristic remark on the periodic 
variation in the number of solar neutrinos detected on Earth. {\it Astrophysics
and Space Science}, {\bf 228}, 113-134.\par
\smallskip
\noindent
Haubold, H.J. and Mathai, A.M. (1998). An integral arising frequently in astronomy and physics. {\it SIAM Review}, {\bf 40}, 995-997.\par
\smallskip
\noindent
Haubold, H.J. and Mathai, A.M. (2000). The fractional kinetic equation and 
thermonuclear functions. {\it Astrophysics and Space Science}, {\bf 273}, 
53-63.\par
\smallskip
\noindent
Honerkamp, J. (1994). {\it Stochastic Dynamical Systems: Concepts, Numerical Methods, Data Analysis}. VCH Publishers, New York.\par
\smallskip
\noindent
Kerridge, D.F. (1961). Inaccuracy and inference. {\it Journal of the Royal 
Statistical Society, Series B}, {\bf 23}, 184-194.\par
\smallskip
\noindent
Mathai, A.M. (1993a). {\it A Handbook of Generalized Special Functions for Statistical and Physical Sciences}, Clarendon Press, Oxford.\par
\smallskip
\noindent
Mathai, A.M. (1993b). The residual effect of a growth-decay mechanism and the 
distributions of covariance structures. {\it The Canadian Journal of Statistics},
{\bf 21(3)}, 277-283.\par
\smallskip
\noindent
Mathai, A.M. (2005). A pathway to matrix-variate gamma and normal densities. {\it 
Linear Algebra and Its Applications}, {\bf 396}, 317-328.\par
\smallskip
\noindent
Mathai, A.M. and Rathie, P.N. (1975). {\it Basic Concepts in Information Theory 
and Statistics: Axiomatic Foundations and Applications}, Wiley Halsted, New York
and Wiley Eastern, New Delhi.\par
\smallskip
\noindent
Mathai, A.M. and Rathie, P.N. (1976). Recent contributions to axiomatic definitions of information and statistical measures through functional equations. In {\it Essays in Probability and Statistics} (eds. S. Ikeda et al.), Shinko Tsusho Co. Ltd., Tokyo, Japan, pp. 607-633.\par
\smallskip
\noindent
Mathai, A.M., Saxena, R.K. and Haubold, H.J. (2005). A certain class of Laplace 
transforms with applications to reaction and reaction-diffusion equations. {\it arXiv.
math.CA/0604472 v2}.\par
\smallskip
\noindent
Saxena, R.K., Mathai, A.M., and Haubold, H.J. (2004a). On generalized fractional kinetic equations. {\it Physica}, {\bf A 344}, 657-664.\par
\smallskip
\noindent
Saxena, R.K., Mathai, A.M., and Haubold, H.J. (2004b). Astrophysical thermonuclear functions for Boltzmann-Gibbs statistics and Tsallis statistics. {\it Physica}, {\bf A 344}, 649-656.\par
\smallskip
\noindent
Tsallis, C. (1988). Possible generalization of Boltzmann-Gibbs statistics. {\it J. Stat. Phys.}, {\bf 52}, 479-487.\par
\smallskip
\noindent
Tsallis, C. (2004). What should a statistical mechanics satisfy to reflect 
nature? {\it Physica D}, {\bf 193}, 3-34.\par
\clearpage
\begin{center}
{\bf Appendix}
\end{center}
\smallskip
\noindent
{\bf A1. Evaluation of the generalized entropy for some standard distributions}
\smallskip
\noindent

Gamma density:
$$f(x)=\frac{x^{\gamma-1}e^{-x/\beta}}{\beta^\gamma\Gamma(\gamma)},$$
for $x>0,\;\gamma >0,\;\beta >0$ and $f(x)=0$ elsewhere. This includes the chisquare
and exponential densities also.
\begin{eqnarray*}
\int_0^\infty[f(x)]^{2-\alpha}dx&=& \frac{\int_0^\infty x^{(\gamma-1)(2-\alpha)}e^{-\frac{(2-\alpha)x}{\beta}}dx}
{\beta^{(2-\alpha)\gamma}[\Gamma(\gamma)]^{2-\alpha}}\nonumber\\
&=&\frac{\beta^{\alpha-1}\Gamma[(\gamma-1)(2-\alpha)+1]}
{[\Gamma(\gamma)]^{2-\alpha}}
\end{eqnarray*}
\noindent
Hence
\begin{eqnarray*}
M_\alpha(f)&=&\frac{1}{\alpha-1}\left[\frac{\beta^{\alpha-1}
\Gamma[(\gamma-1)(2-\alpha)+1]}{[\Gamma(\gamma)]^{2-\alpha}}-1\right]\\
\mbox{and}\\
\lim_{\alpha\rightarrow 1}M_{\alpha}(f)&=&\ln\beta+(1-\gamma)\psi(\gamma)
-\ln\Gamma(\gamma)\\
&=&\ln\beta\; \mbox{for}\;\gamma=\alpha, 
\end{eqnarray*}
where $\psi(\cdot)$ is a psi function or digamma function. It may be also noted 
that the continuous analogue of $M_{k,\alpha}^{*}(P)$ of $(9)$ is
$$M_\alpha^{*}(f)=\frac{\ln(\int_{-\infty}^\infty[f(x)]^{2-\alpha}dx)}{\alpha-1},\; \alpha\neq 1,\; \alpha <2.
$$
For the gamma density $M_\alpha^{*}(f)$ also gives the same quantity as 
above, and $\ln\beta$ when $\alpha\rightarrow 1$.\\
It may be observed that the same procedure goes through even for the pathway 
model of Mathai (2005), in the scalar case, and $M_{\alpha}(f)$ as well as 
$M_{\alpha}^{*}(f)$ can be computed explicitly without much difficulty. The 
pathway model contains a large number of distributions of common use in various 
disciplines. But Shannon entropy $S_k$ of $(6)$ is quite difficult to evaluate, 
in the continuous analogues, even for many of the standard densities in common 
use due to the presence of $\ln f(x)$.
\end{document}